\documentclass[12pt]{article}
\usepackage[dvips]{color}
\usepackage{epsfig}
\usepackage{amsmath}
\usepackage{graphicx}

\date{}
\begin{document}
\title{\bf $AdS_{5}$ black hole at $\mathcal{N}=2$ supergravity}
\author{B Pourhassan$^{1}$\thanks{Email: b.pourhassan@umz.ac.ir}\hspace{1mm}, J Sadeghi$^{2}$\hspace{1mm}and A Chatrabhuti$^{3}$\\
$^1$ {\small {\em  Department of Physics, Imam Hossein University, Tehran, Iran}}\\
$^2$ {\small {\em  Islamic Azad University, Ayatollah Amoli Branch, Amol, Iran}}\\
$^3$ {\small {\em  Theoretical High-energy Physics and Cosmology group, Department of Physics,}}\\
{\small {\em Faculty of Science, Chulalongkorn University, Bangkok
10330, Thailand}}}
\maketitle
\begin{abstract}
\noindent In this paper, we consider the charged non-extremal black
hole at five dimensional $\mathcal{N}=2$ supergravity. We study
thermodynamics of $AdS_{5}$ black hole with three equal charges
($q_{1}=q_{2}=q_{3}=q$). We obtain Schr\"{o}dinger like equation and
discuss the effective potential. Then, we consider the case of the
perturbed dilaton field background and find presence of odd
coefficients of the wave function. Also we find that the higher
derivative corrections have no effect on the first and second even
coefficients of the wave function.
{\bf Pacs:} 04.70.Bw; 11.25.-w; 11.25.Tq
\end{abstract}
\section{. Introduction}
The AdS/CFT correspondence is a powerful mathematical tool to study
the strong coupled gauge theories [1-5]. The AdS/CFT correspondence
or gauge/gravity duality relates a classical gravity on the AdS
space to a gauge theory on the boundary of AdS space. The famous
example of this correspondence is the relation between type IIB
string theory in $AdS_{5}\times S^{5}$ space and $\mathcal{N}=4$
super Yang-Mills gauge theory on the 4-dimensional boundary of
$AdS_{5}$ space. Also recent studies have shown that
there is a relation between the string theory and gauge theory with supersymmetry less than 4 [6-10]. Therefore the subject of supersymmetry [11] is an important point of view.\\
In this paper we consider $AdS_{5}$ black hole at the
$\mathcal{N}=2$ supergravity background and discuss the massless
scalar field in black hole. $\mathcal{N}=2$ supergravity theory is
dual to the $\mathcal{N}=4$ SYM theory with finite chemical
potential [12, 13]. In the recent works, it is found that
$\mathcal{N}=2$ supergravity is an ideal laboratory [14-16].
Application of the AdS/CFT correspondence in the $\mathcal{N}=2$
supergravity background have also been studied [8-10]. By using the
retarded Green's function one can find quasi-normal modes of the
massless scalar field in the black hole. We know that the imaginary
part of the Green's function is used to obtain the quasi-normal
modes [17]. In order to find the quasi-normal modes various methods
have been used [18-35]. For the first time in this paper, we
consider the problem of charged non-extremal $AdS_{5}$ black hole at
the $\mathcal{N}=2$ supergravity background and study various
thermodynamical aspect of this theory. Recently the quasi-normal
modes of the massless scalar field in AdS black hole and scalar
glueballs in a holographic AdS/QCD model at finite temperature have
been considered [34]. This is known as the soft-wall model [36-38].
The vector meson spectrum [36], $m^{2}=4c(n+1)$, scalar glueballs
spectrum, $m^{2}=4c(n+2)$, and vector glueballs spectrum,
$m^{2}=4c(n+3)$, have been discussed, where $n=0, 1, 2,...$ is the
radial quantum number and the parameter $\sqrt{c}$ plays role of a
mass scale [37]. The glueballs are described by massless scalar
field in black hole background of dilaton field $\Phi(r)$. According
to the Maldacena dictionary, the spectrum of scalar glueballs on
boundary are corresponding to quasi-normal modes of the massless
scalar fields in black hole. The $AdS_{5}$ space in Poincare
coordinates is given by the following line element,
\begin{equation}\label{s1}
ds^{2}_{AdS}=\frac{r^{2}}{L^{2}}\left[-dt^{2}+\sum_{i=1}^{3}(dx^{i})^{2}+dr^{2}\right],
\end{equation}
where the constant $L$ denotes radius of curvature of AdS space. If
we include the dilaton field $\Phi(r)$ into the bulk, then the
action extended to the following,
\begin{equation}\label{s2}
S=-\frac{N_{f}}{4N_{c}\pi^{2}}\int{d^{5}x\sqrt{-g}e^{-\Phi}\mathcal{L}},
\end{equation}
where the factor $e^{-\Phi}\mathcal{L}$ denotes interaction of
dilaton field with some matter. So, in absence of some matter the
lagrangian $\mathcal{L}$ do not couple to the dilaton field. In this
paper we will consider the case without matter.
\section{. $\mathcal{N}=2$ supergravity}
In this section we consider a soft-wall model with the dilaton field background $\Phi(r)=cr^{2}$ and the non-extremal $AdS_{5}$ black hole of
$\mathcal{N}=2$ supergravity,
\begin{eqnarray}\label{s3}
ds^{2}&=&-\frac{f}{H^{2}}dt^{2}+H(r^{2} d\Omega_3^{2}+\frac{dr^{2}}{f}),\nonumber\\
f&=&1-\frac{\eta}{r^{2}}+g^{2}r^{2}H^{3} ,\nonumber\\
H&=&1+\frac{q}{r^{2}},
\end{eqnarray}
where $g$ is the coupling constant and relates to the cosmological
constant [43] via $\Lambda=-6g^{2}$, $q$ is called the black hole
charge and is related to non-extremality parameter $\eta$ by using
the relation $q = \eta\sinh^2 \beta$ [8-10]. $\beta$ parameter is
related to the electrical charge of black hole. We note that, in the
$q\rightarrow0$ ($\eta\rightarrow0$) limit, the line element given
by Eq. (3) reduced to line element of an extremal black hole (near
extremal) with zero charge (infinitesimal charge). The coordinate
$r$ is axis along the black hole and is defined in the interval
$0\leq r\leq\infty$. Here, we assume that black hole horizon is
located at $r=r_{h}$ and space-time boundary located at
$r\rightarrow\infty$. Relation between the Hawking temperature
$T_{H}$, horizon radius and black hole charge is given by [44],
\begin{equation}\label{s4}
T_{H}=\frac{2+3k-k^{3}}{2(1+k)^{\frac{3}{2}}}\frac{r_{h}}{\pi
L^{2}},
\end{equation}
where $k\equiv q/r_{h}^{2}$ and radius $L$ relates to the coupling
constant via $g=\frac{1}{L}$. It is clear that the $q\rightarrow0$
($\eta\rightarrow0$) limit of Eq. (4) reduces to the Hawking
temperature of $\mathcal{N}=4$ SYM theory in 4 dimensions, which is
given by $T=r_{h}/\pi L^{2}$ [38]. Moreover, one can see that the
zero temperature limit obtained by taking $r_{h}^2=q/2$ in Eq. (4).
The $\mathcal{N}=2$ $AdS_{5}$ supergravity solution given by Eq. (3)
is dual to the $\mathcal{N}=4$ SYM with finite chemical potential in
Minkowski space. It can be shown by following re-scaling [44],
\begin{equation}\label{s5}
r\rightarrow\lambda^{\frac{1}{4}}r, \hspace{5mm}
t\rightarrow\frac{t}{\lambda^{\frac{1}{4}}}, \hspace{5mm}
\eta\rightarrow\lambda\eta, \hspace{5mm}
q\rightarrow\lambda^{\frac{1}{2}}q,
\end{equation}
and taking $\lambda\rightarrow\infty$ limit while,
\begin{equation}\label{s6}
d\Omega_{3}^{2}\rightarrow\frac{1}{L^{2}\lambda^{\frac{1}{2}}}(dx^{2}+dy^{2}+dz^{2}),
\end{equation}
and also we set $r_{0}^{4}\equiv\eta R^{2}$. Then solution given by
Eq. (3) reduces to the following,
\begin{eqnarray}\label{s7}
ds^{2}&=&e^{2A(r)}\left[-\frac{f}{H^{2}}dt^{2}+H d{\vec{X}}^{2}+\frac{H}{f}dr^{2}\right],\nonumber\\
f&=&H^{3}-\frac{r_{0}^{4}}{r^{4}} ,\nonumber\\
H&=&1+\frac{q}{r^{2}},
\end{eqnarray}
where the geometric function $A(r)$ is defined as
$A(r)\equiv\ln{\frac{r}{L}}$. The coordinate $r$ is defined in the
interval $r_{0}\leq r\leq\infty$, and $r=\infty$ corresponds to the
space boundary. It is interesting to consider special case of one
charge black hole ($q_{1}=q, q_{2}=q_{3}=0$). In that case the
Hawking temperature reads as,
\begin{equation}\label{s8}
T_{H}=\frac{q+2r_{0}^{2}}{2\pi L^{2}\sqrt{q+r_{0}^{2}}}.
\end{equation}
The radius of the horizon is given by,
\begin{equation}\label{s9}
r_{h}^{2}=\frac{1}{2}\left(\sqrt{4r_{0}^{4}+q^{2}}-q\right).
\end{equation}
Also one can write the chemical potential in terms of the black hole
charge and horizon radius,
\begin{equation}\label{s10}
\mu=\frac{r_{h}}{L^{2}}\sqrt{\frac{2q}{q+r_{h}^{2}}}.
\end{equation}
Therefore, at $q=0$ limit is equal to the zero chemical potential
limit. Approximately, one can say that the $\sqrt{-g}$ in the action
given by Eq. (2) is same for both $q\neq0$ and $q=0$ cases. The case
of $q=0$ is studied by Miranda et al. [34], but for the charged
black hole one can obtain,
\begin{equation}\label{s11}
r_{0}^{2}=\frac{\pi^{2}T_{H}^{2}L^{4}-q}{2}\left[1+\sqrt{1+q\frac{4\pi^{2}T_{H}^{2}L^{4}-q}{(\pi^{2}T_{H}^{2}L^{2}-q)^{2}}}\right].
\end{equation}
Then by using the metrics given by Eqs. (1) and (7) in the action
given by Eq. (2), the $S_{BH}$ obtained in terms of the $r_{0}$. The
specific heat is important parameter to find the phase transition
which will be calculated by relation
${\mathcal{C}}_{BH}\equiv-\beta^{2}\frac{\partial^{2}}{\partial\beta^{2}}S_{BH}$.
If sign of the specific heat is positive (negative), black hole will
be in a metastable (unstable) phase [34]. In that case one can
obtain,
\begin{equation}\label{s12}
{\mathcal{C}}_{BH}
\propto\frac{{\tilde{T}}^{2}}{\sqrt{{\tilde{T}}^{2}-q}}\left[(6c-9q+\frac{2c{\tilde{T}}^{2}}{{\tilde{T}}^{2}-q}+12{\tilde{T}}^{2})
e^{-\frac{c}{{\tilde{T}}^{2}-q}}-6{\tilde{T}}^{2}\right]+\mathcal{O}(q^{2}),
\end{equation}
where $\tilde{T}=\pi L^{2} T_{H}$ and $\beta=\frac{1}{T_{H}}$. This
is clear that the case $q=0$ recovers results given by Miranda et
al. [34]. In the $c\rightarrow0$ limit the sign of specific heat is
positive for $q=0$, but in our case with $q\neq0$ the sign of
specific heat depends on black hole charge, so if
${\tilde{T}}^{2}>1.5 q$, then the charged black hole is in stable
phase. In [34] it is found that the specific heat changes sign at
${\tilde{T}}^{2}\simeq0.75$ (for $c\neq0$ and $q=0$). In presence of
the dilaton field ($c\neq0$) and in unit of $c$ one can find that
the phase transition temperature increases in case of charged black hole.
For example, in case of $q=1$ we find that phase transition happens at ${\tilde{T}}^{2}\simeq2.4$, so the charged black hole is in stable phase for ${\tilde{T}}^{2}>2.4$.\\
A similar study was considered by using the original metric given by
Eq. (3) by Sadeghi et al. [45]. In this paper we consider re-scaled
metric given by Eq. (7).
\section{. Equation of motion}
As we know, the scalar glueball on boundary of space-time is dual of
massless scalar field in black hole. Now, we solve the equation of
motion for scalar field in black hole background and obtain wave
functions. In order to obtain wave functions we rewrite the equation
of motion in Schr\"{o}dinger like equation. Then by using the
asymptotic behavior of Schr\"{o}dinger like equation, we obtain
solutions of this equation. The massless scalar field of this theory
is described by the following action,
\begin{equation}\label{s13}
S=-\frac{\pi^{3}L^{5}}{4\kappa_{10}^{2}}\int{d^{5}x\sqrt{-g}e^{-\Phi}g^{MN}\partial_{M}\phi\partial_{N}\phi},
\end{equation}
where $\Phi(r)=cr^{2}$ is dilaton field and $g_{MN}$ is the metric
which is given by Eq. (7). We note that $\mathcal{N}=2$ supergravity
in presence of dilaton field has been already studied [46, 47]. The
parameter $\kappa_{10}$ is ten-dimensional gravity constant. Indices
$M$ and $N$ run from $0$ to $4$, so coordinates $x^{\mu}$
($\mu=0,1,2,3$) describe four-dimensional boundary and $x^{4}=r$ is
extra coordinate along the black hole. By using equation of motion
for the scalar field $\phi$ in Eq. (13) one can obtain,
\begin{equation}\label{s14}
\frac{e^{\Phi}}{\sqrt{-g}}\partial_{r}
(\sqrt{-g}e^{-\Phi}g^{rr}\partial_{r}\phi)+g^{\mu\nu}\partial_{\mu}\partial_{\nu}\phi=0.
\end{equation}
Now, we use Fourier transformation,
\begin{equation}\label{s15}
\phi(r,x)=\int{\frac{d^{4}k}{(2\pi)^{4}}e^{ik.x}\bar{\phi}(r,k)}.
\end{equation}
We put $k_{\mu}=(-\omega, p_{i})$, to Eq. (14) and obtain,
\begin{equation}\label{s16}
f(r)e^{D}\partial_{r} \left(e^{-D}f(r)\partial_{r}\bar{\phi}\right)
+\left(H^{3}\omega^{2}-p^{2}f(r)\right)\bar{\phi}=0,
\end{equation}
where we set $p^{2}=\Sigma_{i-1}^{3} p_{i}^{2}$ and $D\equiv
cr^{2}-3A(r)$. Now, we introduce new coordinates
$\partial_{r_{\ast}}=-f(r)\partial_{r}$. These are well known as
Regge-Wheeler coordinates, so we can integrate and find explicit
expression of $r_{\ast}$ in terms of $r$,
\begin{equation}\label{s17}
r_{\ast}=r+\frac{1}{B-A}\left[A^{\frac{3}{2}}\tan^{-1}{\frac{r}{\sqrt{A}}}-B^{\frac{3}{2}}\tan^{-1}{\frac{r}{\sqrt{B}}}\right],
\end{equation}
where $B\equiv\frac{1}{2}(q+\sqrt{q^{2}+4r_{0}^{4}})$ and
$A=-\frac{r_{0}^{4}}{B}$. It is expected that in $q=0$ limit our
results are same as results given by Miranda et al. [34], but we
note that our coordinates is different with that paper, so any
difference at $q=0$ limit is natural. Later, we discuss how two
results coincide. Then we choose new variable as
$\psi=e^{-\frac{D}{2}}\bar{\phi}$, where $D\equiv
cr^{2}-3\ln{\frac{r}{L}}$. By using above definitions in Eq. (16)
one can obtain the following Schr\"{o}dinger like equation,
\begin{equation}\label{s18}
\partial_{r_{\ast}}^{2}\psi+\omega^{2}\psi=V\psi,
\end{equation}
where we define the effective potential as following,
\begin{equation}\label{s19}
V=p^{2}f(r)-\frac{D^{\prime\prime}}{2}+\frac{D^{\prime2}}{4}+\mathcal{O}(q).
\end{equation}
In the effective potential given by Eq. (19), prime denotes the
derivative with respect ro the $r_{\ast}$ and
$\mathcal{O}(q)=-(\frac{3q}{r^{2}}+\frac{3q^{2}}{r^{4}}+\frac{q^{3}}{r^{6}})\omega^{2}$.
The explicit expression of the effective potential in terms of $r$
written as following relation,
\begin{equation}\label{s20}
V=\frac{f(r)}{r^{2}}\left[p^{2}r^{2}+2qc-\frac{3q}{r^{2}}-4c\frac{r_{0}^{4}}{r^{2}}+6\frac{r_{0}^{4}}{r^{4}}+f(r)
(c^{2}r^{4}+\frac{3}{4}-3c-cr^{2})\right].
\end{equation}
In order to compare our results with work of Miranda et al. [34] we
change our coordinates. So in the new coordinates we have
$f(z)=1+qz^{2}-(\frac{z}{z_{0}})^{4}$ ($z_{0}\sim\frac{1}{r_{0}}$)
and $D=cz^{2}+3\ln{\frac{z}{L}}$, so the effective potential is read
as,
\begin{equation}\label{s21}
V=\frac{f(z)}{z^{2}}\left[p^{2}z^{2}+\frac{15}{4}+\frac{9}{4}(\frac{z}{z_{h}})^{4}+2cz^{2}(1+(\frac{z}{z_{h}})^{4})
+c^{2}z^{4}f(z)+\frac{3}{4}qz^{2}\right].
\end{equation}
In Fig. 1 we plot the effective potential given by Eq. (21) in terms of the radial coordinate $z$ for $q=1$.\\
Corresponding to the asymptotical behavior of equation of motion of
the scalar field one can write two solutions which satisfy the
Schr\"{o}dinger like equation,
\begin{eqnarray}\label{s22}
\psi_{1}&=&z^{\frac{5}{2}}[1+a_{1}z+a_{2}z^{2}+a_{3}z^{3}+a_{4}z^{4}+\ldots]\nonumber\\
\psi_{2}&=&z^{-\frac{3}{2}}[1+b_{1}z+b_{2}z^{2}+b_{3}z^{3}+\ldots]+b_{4}\psi_{1}\ln(cz^{2}).
\end{eqnarray}
Only none-zero coefficients are $a_{2}$, $a_{4}$, $b_{2}$ and
$b_{4}$. One can interpret $\psi_{2}-b_{4}\psi_{1}\ln(cr^{2})$ as a
source of an operator on the boundary. On the other hand we know
that glueball operator is dual of massless scalar field in to the
bulk [36]. The non-zero coefficients of the wave function given by
Eq. (22) are given by,
\begin{eqnarray}\label{s23}
a_{2}&=&-\frac{\omega^{2}-p^{2}-2c+8q}
{12},\nonumber\\
a_{4}&=&\frac{(\omega^{2}-p^{2}-2c+8q)^{2}}{384}+\frac{1}{2z_{h}^{4}}+\frac{c^{2}}{32}+\frac{q}{32}(p^{2}+2c-8q),\nonumber\\
b_{2}&=&\frac{\omega^{2}-p^{2}-2c+8q}{4}.
\end{eqnarray}
These coefficients recover previous results [34] at $q=0$.
We use above solutions to construct retarded Green's function and then one can extract quasi-normal modes.\\
One can see that at $z\rightarrow z_{h}$
($z_{h}^4=\frac{z_{0}^{4}}{4}(qz_{0}^{2}+\sqrt{4+q^{2}z_{0}^{4}})^{2}$),
the effective potential vanishes. Therefore, Schr\"{o}dinger like
equation has other solutions as incoming and outgoing wave functions
are denoted by $\psi_{+}$ and $\psi_{-}$. These solutions are
obtained by using the fact that effective potential vanishes at the
horizon and the Schr\"{o}dinger like equation takes the form of free
particle equation with solution of $exp(\pm i\omega r_{\ast})$. The
negative sign is interpreted as incoming plane wave in to the
horizon and the positive sign interpreted as the outgoing plane wave
from the horizon. Therefore, one can write the following expansion
as the near horizon solution of the Schr\"{o}dinger like equation,
\begin{equation}\label{s24}
\psi_{\pm}=e^{\pm i\omega
r_{\ast}}\left[1+a_{1(\pm)}(1-\frac{z}{z_{h}})+a_{2(\pm)}(1-\frac{z}{z_{h}})^{2}+\cdots\right].
\end{equation}
$\psi_{\pm}$ form the basis of any other wave functions and may be
expanded in terms of $\psi_{1}$ and $\psi_{2}$. Therefore, one can
relate solutions given by Eq. (22) to (24) and vise versa,
\begin{eqnarray}\label{s25}
\psi_{\pm}&=&{\mathcal{A}}_{(\pm)}\psi_{2}+{\mathcal{B}}_{(\pm)}\psi_{1},\nonumber\\
\psi_{2,1}&=&{\mathcal{C}}_{(2,1)}\psi_{-}+{\mathcal{D}}_{(2,1)}\psi_{+},
\end{eqnarray}
where ${\mathcal{A}}_{(+)}$, ${\mathcal{A}}_{(-)}$,
${\mathcal{B}}_{(+)}$, ${\mathcal{B}}_{(-)}$, ${\mathcal{C}}_{(2)}$,
${\mathcal{C}}_{(1)}$, ${\mathcal{D}}_{(2)}$ and
${\mathcal{D}}_{(1)}$ are $\omega$ and $p$-dependent coefficients,
which are determined by using boundary conditions and are related to
each other by the following relation,
\begin{eqnarray}\label{s26}
1=\left(\begin{array}{ccc}
{\mathcal{A}}_{(-)}&{\mathcal{B}}_{(-)}\\
{\mathcal{A}}_{(+)}&{\mathcal{B}}_{(+)}\\
\end{array}\right)\left(\begin{array}{ccc}
{\mathcal{C}}_{(2)}&{\mathcal{D}}_{(2)}\\
{\mathcal{C}}_{(1)}&{\mathcal{D}}_{(1)}\\
\end{array}\right).
\end{eqnarray}
This relation will be useful to find quasi-normal modes.\\
Now, we discuss the effective potential at special limits. Here, we
set $p=0$ and $c=0$, which is corresponding to case of the charged
black hole without dilaton field. Therefore, the effective potential
is reduced to following expression,
\begin{equation}\label{s27}
V=\frac{f(z)}{z^{2}}\left[\frac{15}{4}+\frac{9}{4}(\frac{z}{z_{h}})^{4}+\frac{3}{4}qz^{2}\right].
\end{equation}
It is clear that the effective potential is enhanced due to black
hole charge.
\section{. Imaginary part of the retarded Green's function}
The retarded Green's function helps us to obtain the quasi-normal
modes of scalar field $\phi$ in the black hole background. It is
known that the imaginary part of the retarded Green's function is
called the spectral function which is used to find the frequency of
the quasi-normal modes. In this paper we use previous results [34]
to write the retarded Green's function. So, one can rewrite the Eq.
(13) as,
\begin{equation}\label{s28}
S=-\frac{\pi^{3}L^{5}}{4\kappa_{10}^{2}}\int{d^{4}xdz\partial_{z}(\frac{f}{H}e^{-2A(z)}
\sqrt{-g}e^{-\Phi}\phi\partial_{z}\phi)},
\end{equation}
where we used equation of motion and $\phi_{k}(z)$ satisfies this
equation. So, one can expand it in terms of obtained solutions.
Hence, for case of finite temperature one can obtain,
\begin{equation}\label{s29}
\phi_{k}(z)=e^{\frac{D}{2}}\left[\psi_{2}(z)+\frac{\mathcal{B}_{(-)}}{\mathcal{A}_{(-)}}\psi_{1}(z)\right].
\end{equation}
In Eq. (28) we integrate over $z$ and set $\phi(z_{h})=0$. Then, it
is found that imaginary part of the retarded Green's function is
proportional to $\mathcal{B}_{(-)}/\mathcal{A}_{(-)}$ which is
called spectral function. In AdS/CFT correspondence, the spectral
function is in AdS side which is dual of the correlation function in
the other side. Therefore, we can use following relation [36] to
obtain imaginary part of the retarded Green's function,
\begin{equation}\label{s30}
G\equiv\frac{\mathcal{B}_{(-)}}{\mathcal{A}_{(-)}}=\frac{\partial_{z}\psi_{-}\psi_{2}-\psi_{-}\partial_{z}\psi_{2}}
{\partial_{z}\psi_{-}\psi_{1}-\psi_{-}\partial_{z}\psi_{1}}.
\end{equation}
In Eq. (30) we evaluate functions in the near horizon limit. In that
case, one can obtain,
\begin{equation}\label{s31}
ImG=\omega\frac{\psi_{2}\partial_{z}\psi_{1}-\psi_{1}\partial_{z}\psi_{2}} {(\omega\psi_{1})^{2}+(\partial_{z}\psi_{1})^{2}},
\end{equation}
where $\psi_{1}$ and $\psi_{2}$ are given by Eq. (22) at
$z\rightarrow z_{h}$ limit. Then, one can find the value of Eq. (31)
numerically. In that case we draw zero-momentum case of imaginary
part of retarded Green's function (spectral functions) in terms of
temperature for various values of  frequency as shown in Fig. 2.\\
There are different ways to obtain the quasi-normal modes in black
holes, such as power series method [48] which is suitable for
asymptotic AdS space-time. In this case we begin with the
Schr\"{o}dinger like equation given by Eq. (18). We introduce new
function $\psi=\varphi e^{-i\omega r_{\ast}}$ and use Eq. (18) to
obtain,
\begin{equation}\label{s32}
\partial_{r_{\ast}}^{2}\varphi-2i\omega\partial_{r_{\ast}}\varphi=V\varphi.
\end{equation}
Now, by using the relation $\partial_{r_{\ast}}=-f(z)\partial z$, we
obtain following equation,
\begin{equation}\label{s33}
f(z)\frac{\partial^{2}\varphi}{\partial
z^{2}}+\left(2i\omega+2qz-\frac{4z^{3}}{z_{h}^{4}}\right)\frac{\partial\varphi}{\partial
z}-\frac{V(z)}{f(z)}\varphi=0.
\end{equation}
Because of a regular singularity at $z=z_{h}$ one can write a
possible solution of Eq. (33) as,
\begin{equation}\label{s34}
\varphi_{-}(r)=\sum_{m}a_{m(-)}(1-\frac{z}{z_{h}})^{m}.
\end{equation}
In order to find coefficients $a_{m(-)}$ we should put $\varphi_{-}$
into the equation (33), then by using the Dirichlet boundary
condition we have,
\begin{equation}\label{s35}
\sum_{m}a_{m(-)}=0.
\end{equation}
This leads us to obtain the roots of Eq. (35) which may be performed
numerically, then quasi-normal frequencies can be obtained.
\section{. Perturbed dilaton field background}
The light scalar and vector glueballs spectra in a holographic QCD
with a perturbed dilaton field background originally have studied in
[36]. AdS/QCD duality in a perturbed AdS-dilaton background was
considered [37]. In the mentioned works, the equation of motion for
bulk-to-boundary propagator and AdS two-point correlation function
have been obtained and concluded that the AdS dynamics is not
affected by dilaton perturbations. Now we would like to apply
perturbed dilaton background to Schr\"{o}dinger like equation and
obtain effect of perturbation on effective potential and
solutions of wave function.\\
It is known that the dilaton field background and geometric function
should satisfy the Regge behavior. The simplest choice consistent
with these conditions are given in previous section. Also there is
another choice for dilaton field background $\Phi(z)$, which obeys
the Regge behavior. One can consider perturbed dilaton field
background as following [36],
\begin{eqnarray}\label{s36}
\Phi(z)&=&cz^{2}+\sqrt{c}\lambda z\nonumber\\
A(z)&=&-\ln \frac{z}{L},
\end{eqnarray}
where $\lambda$ is perturbation parameter which is small
dimensionless parameter. The choice of Eq. (36) does not effectively
modify Regge behavior of the spectrum. It is found that for small
values of the parameter $\lambda$, first three spectra of scalar
glueball can be modified as,
\begin{eqnarray}\label{s37}
m_{0}^{2}&=&8c+\lambda\frac{3\sqrt{\pi}}{2},\nonumber\\
m_{1}^{2}&=&12c+\lambda\frac{27\sqrt{\pi}}{16},\nonumber\\
m_{2}^{2}&=&16c+\lambda\frac{237\sqrt{\pi}}{128}.
\end{eqnarray}
Also, first three spectra of vector glueball can be modified as,
\begin{eqnarray}\label{s38}
m_{0}^{2}&=&12c+\lambda\frac{189\sqrt{\pi}}{128},\nonumber\\
m_{1}^{2}&=&16c+\lambda\frac{105\sqrt{\pi}}{64},\nonumber\\
m_{2}^{2}&=&20c+\lambda\frac{14667\sqrt{\pi}}{8192}.
\end{eqnarray}
It has been shown that mass splitting between vector and scalar
qlueballs increases if $\lambda$ is negative. Now we would like to
obtain the modified waves functions and the effective potential
corresponding to the Schr\"{o}dinger like equation.\\
By using perturbed dilaton background field given by Eq. (36) one
can also obtain Schr\"{o}dinger like given by Eq. (18) where the
effective potential is modified as follows,
\begin{eqnarray}\label{s39}
V&=&\frac{f(z)}{z^{2}}
\left[p^{2}z^{2}+\frac{15}{4}+\frac{9}{4}\frac{z^{4}}{z_{h}^{4}}+2cz^{2}(1+\frac{z^{4}}{z_{h}^{4}})+c^{2}z^{4}f(z)
+\frac{3}{4}qz^{2}\right]\nonumber\\
&+&\frac{f(z)}{z^{2}}\sqrt{c}\lambda
\left[2\frac{z^{5}}{z_{h}^{5}}-qz^{3}+(\frac{\sqrt{c}\lambda
z^{2}}{4}+c z^{3}+\frac{3}{2}z)f(z)\right].
\end{eqnarray}
In Fig. 3 we plot the effective potential given by Eq. (39) in terms
of radial coordinate $z$ for $q=\lambda=1$. We find that dilaton
perturbation increases value of effective potential. An important result is the absence of any potential well so that there are no bound states such as before.\\
Also, coefficients of solution given by Eq. (22) modified as
follows,
\begin{eqnarray}\label{s40}
a_{1}&=&\frac{9\sqrt{c}\lambda}{8},\nonumber\\
a_{2}&=&-\frac{\omega^{2}-p^{2}-2c+8q}{12}+\frac{31}{192}c\lambda^{2},\nonumber\\
a_{3}&=&\frac{\sqrt{c}\lambda}{42}\left[\frac{67}{64}\sqrt{c}\lambda^{2}+\frac{5}{2}p^{2}
+7c-\frac{5}{2}\omega^{2}-\frac{253}{4}q\right],\nonumber\\
a_{4}&=&\frac{(q^{2}+2c-\omega^{2}-8q)^{2}}{384}+\frac{1}{2z_{h}^{4}}+\frac{c^{2}}{32}+q(p^{2}+2c-8q),\nonumber\\
&+&c\lambda^{2}
\left[\frac{11}{8}c+\frac{15}{168}(p^{2}-\omega^{2})-\frac{55}{112}q+\frac{35}{192}(p^{2}+2c-\omega^{2}-8q)\right]+\mathcal{O}(\lambda^{4}),\nonumber\\
b_{1}&=&-\frac{6\sqrt{c}\lambda}{17},\nonumber\\
b_{2}&=&\frac{\omega^{2}-p^{2}-2c+8q}{4}-\frac{3c\lambda^{2}}{4},\nonumber\\
b_{3}&=&\frac{2\sqrt{c}\lambda}{51}\left[15+\frac{99}{4}(\omega^{2}-p^{2}-2c+8q)\right]+\mathcal{O}(\lambda^{3}).
\end{eqnarray}
The main consequence of perturbed dilaton background is appearance
of odd coefficients ($a_{1}$, $a_{3}$, $b_{1}$ and $b_{3}$).\\
Now, we discuss about special behaviors of the modified effective
potential. At zero-temperature limit, where $z_{h}\rightarrow\infty$
and $q=0$, the effective potential takes following form,
\begin{equation}\label{s41}
V_{T=0}(z)=\frac{1}{z^{2}}
\left[(p^{2}+2c)z^{2}+\frac{15}{4}+c^{2}z^{4}+\frac{c\lambda^{2}z^{2}}{4}+\sqrt{c}\lambda
z(\frac{3}{2}+cz^{2})\right].
\end{equation}
Also at $c\rightarrow0$ limit, for finite temperature, one can
obtain,
\begin{equation}\label{s42}
V_{c=0}(z)=\frac{f(z)}{z^{2}}
\left(p^{2}z^{2}+\frac{15}{4}+\frac{9}{4}\frac{z^{4}}{z_{h}^{4}}+\frac{3}{4}qz^{2}\right),
\end{equation}
which corresponds to case without dilaton field. Therefore, in the
above expression the parameter $\lambda$ does not
exist.\\
By increasing the temperature, effective potential changes. It is
known that (for non-perturbed dilaton field background) at
high-temperature there are no bound states. This situation is
similar to recent case with perturbed dilaton field
background [34].\\
At low-temperature limit and $q=0$, one can write effective
potential approximately as,
\begin{eqnarray}\label{s43}
V(z)&=&\frac{1}{z^{2}}
\left[\frac{15}{4}-\frac{3}{2}\frac{z^{4}}{z_{h}^{4}}+2cz^{2}+c^{2}z^{4}-2c^{2}\frac{z^{8}}{z_{h}^{4}}\right]\nonumber\\
&+&\frac{1}{z^{2}}\left[\sqrt{c}\lambda
\left(\frac{3}{2}z+cz^{3}-\frac{3}{2}\frac{z^{4}}{z_{h}^{4}}+\frac{1}{2}\frac{z^{5}}{z_{h}^{4}}-2c\frac{z^{7}}{z_{h}^{4}}\right)\right]\nonumber\\
&+&\frac{1}{z^{2}}\left[c\lambda^{2}
\left(\frac{z^{2}}{4}-\frac{z^{6}}{2z_{h}^{4}}\right)\right].
\end{eqnarray}
Then by using relation $z_{h}=\frac{1}{\pi T}$ one can obtain,
\begin{eqnarray}\label{s44}
V(z)&=&\frac{1}{z^{2}}
\left[\frac{15}{4}+2cz^{2}+c^{2}z^{4}+\sqrt{c}\lambda
z(\frac{3}{2}+cz^{2})+\frac{c\lambda^{2}z^{2}}{4}\right]\nonumber\\
&-&z^{2}\pi^{4}T^{4}\left[\frac{3}{2}+2c^{2}z^{4}-\sqrt{c}\lambda
z(\frac{1}{2}-\frac{3}{2z}-2cz^{2})+\frac{c\lambda^{2}z^{2}}{2}\right].
\end{eqnarray}
The modified effective potential at low-temperature has a constant
term as $(2+\frac{\lambda^{2}}{4})c$, a perturbed oscillator like
potential term as $c^{2}z^{2}+c^{\frac{3}{2}}\lambda z$, a term with
bare infinity at $z=0$ as $\frac{15}{4z^{2}}+\frac{3\sqrt
{c}\lambda}{2z}$, and finally temperature corrections terms. Eq.
(44) agrees with Eq. (43) at $T=0$ limit.
\section{. Higher derivative corrections}
In this section we calculate the effect of higher derivative terms
on the effective potential and wave functions for case of
non-perturbed dilaton field [49,50]. In this way we use results of
[51], where the first-order correction of the solution given by Eq.
(3) is,
\begin{eqnarray}\label{s45}
f&=&1+\frac{r^{2}+q}{L^{2}}-\frac{\eta}{r^{2}}-\frac{1}{r^{4}}(\frac{\alpha}{r^{2}+q}+\beta) ,\nonumber\\
H&=&1+\frac{q}{r^{2}}-\frac{\beta L^{2}}{5r^{4}(r^{2}+q)},
\end{eqnarray}
where,
\begin{eqnarray}\label{s46}
\alpha&=&c_{1}\left(\frac{q(q+\eta)}{72L^{2}}-\frac{\eta^{2}}{96}\right) ,\nonumber\\
\beta&=&\frac{c_{1}q(q+\eta)}{9L^{2}},
\end{eqnarray}
and $c_{1}$ is arbitrary constant. In this background the horizon
radius $r_{h}$ is root of following equation,
\begin{equation}\label{s47}
1+\frac{r^{2}+q}{L^{2}}-\frac{\eta}{r^{2}}-\frac{1}{r^{4}}(\frac{\alpha}{r^{2}+q}+\beta)=0.
\end{equation}
Again we use rescaling of Eq. (5) and take
$\lambda\rightarrow\infty$ limit to get,
\begin{eqnarray}\label{s48}
f&=&1+\frac{q}{r^{2}}-\frac{r_{0}^{4}}{r^{4}}+\frac{c_{1}r_{0}^{8}}{96L^{2}r^{6}(r^{2}+q)} ,\nonumber\\
H&=&1+\frac{q}{r^{2}}-\frac{c_{1}r_{0}^{4}}{9L^{4}r^{4}(r^{2}+q)},
\end{eqnarray}
and now radius $r_{h}$ is root of following relation,
\begin{equation}\label{s49}
r^{8}+2qr^{6}+(q^{2}-r_{0}^{4})r^{4}-qr_{0}^{4}r^{2}+\frac{c_{1}r_{0}^{4}}{L^{2}}=0.
\end{equation}
For special case of $q=0$, one can obtain solution of the above
relation as,
\begin{equation}\label{s50}
r_{h}^{4}=\frac{r_{0}^{4}}{2}\left(1+\sqrt{1-\frac{4c_{1}}{L^{2}}}\right).
\end{equation}
Then similar to section 3 we change our coordinates to have,
\begin{eqnarray}\label{s51}
f&=&1-\frac{z^{4}}{z_{0}^{4}}+\frac{c_{1}z^{8}}{96L^{2}z_{0}^{8}} ,\nonumber\\
H&=&1-\frac{c_{1}z^{6}}{9L^{4}z_{0}^{4}}.
\end{eqnarray}
In this case the effective potential is modified as,
\begin{equation}\label{s52}
V=\frac{f(z)}{z^{2}}\left[p^{2}z^{2}+\frac{15}{4}+\frac{9}{4}(\frac{z}{z_{h}})^{4}+2cz^{2}(1+\frac{z^{4}}{z_{h}^{4}}
-\frac{c_{1}}{32L^{2}}\frac{z^{8}}{z_{h}^{8}})
+c^{2}z^{4}f(z)+\frac{11c_{1}}{128L^{2}}\frac{z^{8}}{z_{h}^{8}}\right],
\end{equation}
where $f(z)$ is given by Eq. (51) and
$z_{h}^{4}=\frac{4Lz_{0}^{4}}{c_{1}}(12L+\sqrt{144L^{2}-6c_{1}})$.
Comparison of Eqs.(21) and (52) reveals that the higher order
correction term is obtained as,
\begin{equation}\label{s53}
-\frac{c_{1}}{32L^{2}}\frac{z^{8}}{z_{h}^{8}}f(z)\left(C+\frac{11}{4z^{2}}\right),
\end{equation}
which decreases the effective potential if
$z^{8}<\frac{z_{h}^{8}}{1-\frac{c_{1}}{96L^2}}$ and increases the
effective potential if
$z^{8}>\frac{z_{h}^{8}}{1-\frac{c_{1}}{96L^2}}$, so at the horizon
the higher order corrections have no effect on the effective
potential. Actually in the case of higher order corrections,
effective potential vanishes at horizon just like all previous
cases. Finally we investigate the effect of higher order corrections
on the wave function given by Eq. (22). In that case it is clear
that the higher order terms have no effect on the first and second
even coefficients, so by setting $q=0$ in Eq. (23) we get same
coefficients. The lowest coefficient, which involves higher
derivative corrections, is $a_{6}$. In Fig. 4 we plot simultaneously
the effective potential for cases of perturbed dilaton, higher
derivative and without them for $q=0$. We observe that there are no
potential wells indicating that there are no bound states.
\section{. Conclusions}
First of all we reviewed $\mathcal{N}=2$ $AdS_{5}$ supergravity
which includes a non-extremal black hole with three equal charges.
We know that $\mathcal{N}=2$ $AdS_{5}$ supergravity solution is dual
of $\mathcal{N}=4$ SYM with finite chemical potential. We used this
duality and calculated the specific heat and found that the black
hole phase transition happens at ${\tilde{T}}^{2}\simeq2.4$.
Comparison of this result with case of the zero charge [34] tells us
that effect of black hole charge increases phase transition
temperature. Then, we obtained wave functions and effective
potential, and found dependence of coefficients and effective
potential on black hole charge. Then we discussed about imaginary
part of retarded Green's function numerically and found that black
hole phase transition temperature is $T=0.1$.\\
The effect of perturbed dilaton field background on wave function
and the effective potential is studied in section 5, we found that
odd coefficients of wave function has been raised. We obtained the
effective potential for the cases of zero-temperature and zero
dilaton field, also obtained case of low-temperature and have shown
that the constant and oscillator like terms were modified by square
and linear form of perturbation parameter respectively. In that case
it is interesting to consider perturbed geometric function [36, 37].
Finally we considered the higher derivative corrections which have
no contribution to the first and second even coefficients, so the
first effect of higher derivative terms is $a_{6}$. The effective
potential is modified by two terms proportional to $z^{8}f(z)$ and
$z^{6}f(z)$. In all cases we observed that there are no potential
wells at finite temperature. We conclude that the dilaton
perturbation increases the effective potential, but higher
derivative correction decreases the effective potential for
$z<z_{h}$.

\begin{figure}[th]
\begin{center}
\includegraphics[scale=.35]{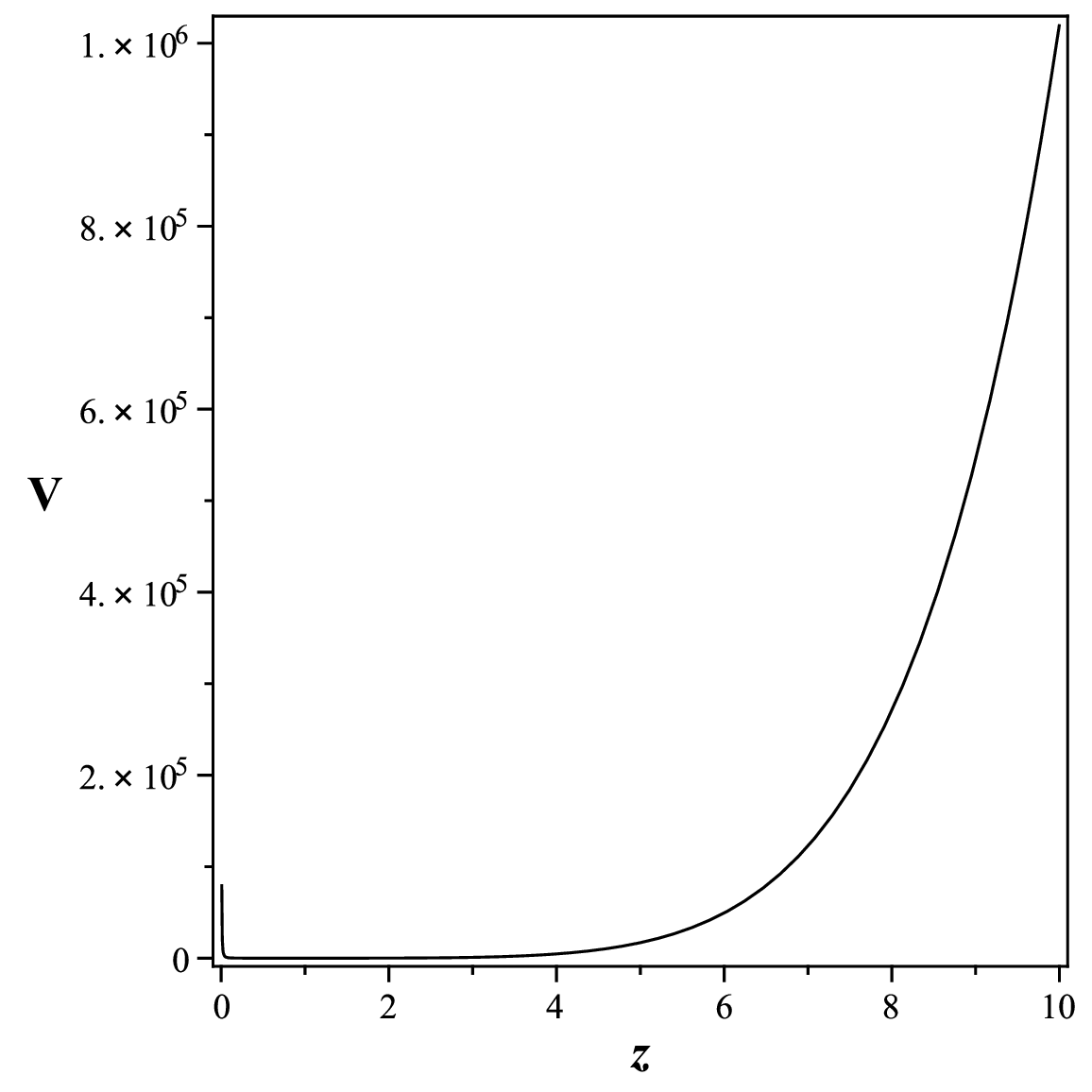}
\caption{Plot of effective potential as a function of $z$ for
$q=c=p=1$.}
\end{center}
\end{figure}

\begin{figure}[th]
\begin{center}
\includegraphics[scale=.4]{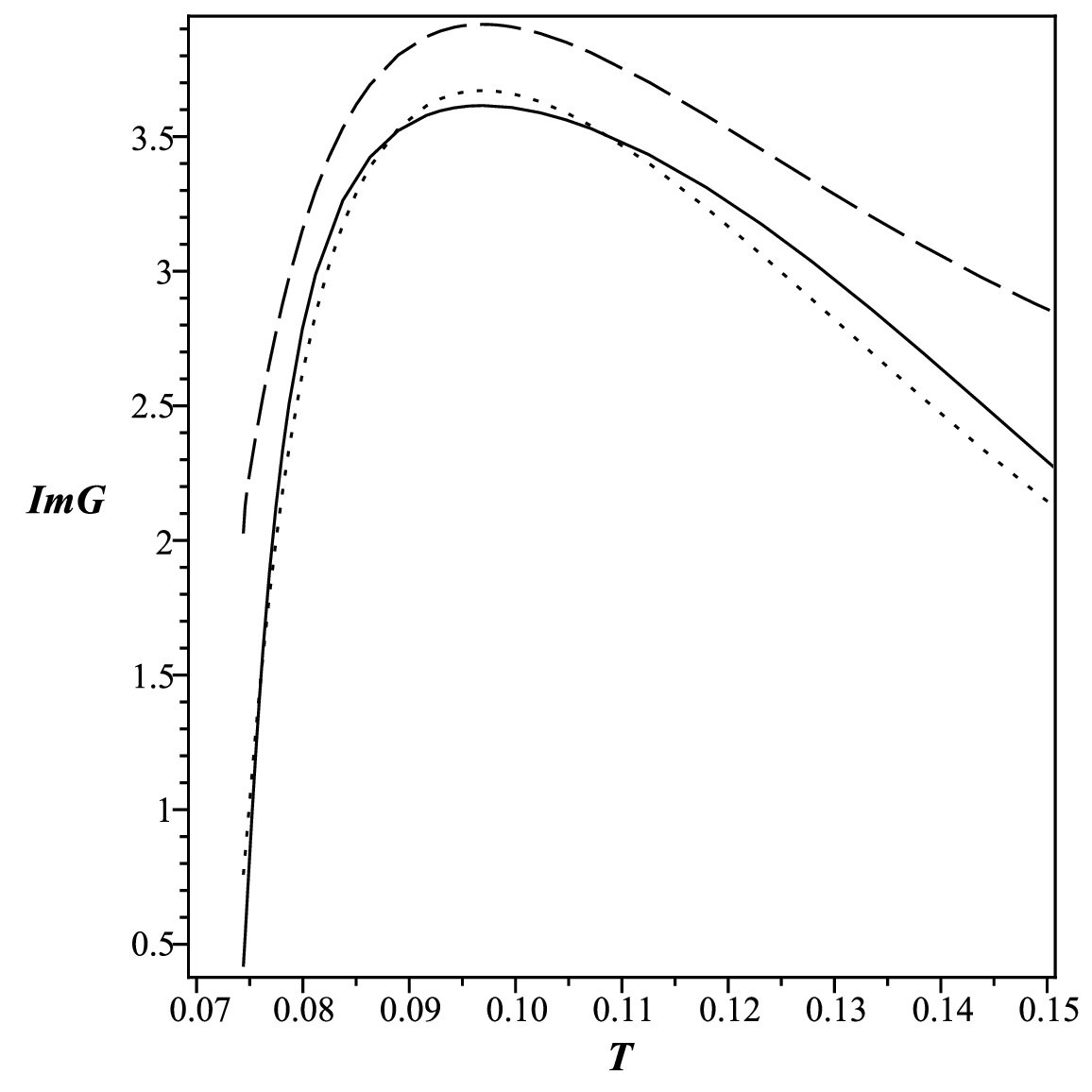}
\caption{Spectral functions plotted as a function of temperature for
$p=0$, $c=q=1$. Solid, dotted and dashed lines are obtained by
choosing $\omega=3.5$, $4$, and $5$ respectively. The peak of
spectral functions show phase transition temperature is $T=0.1$.}
\end{center}
\end{figure}

\begin{figure}[th]
\begin{center}
\includegraphics[scale=.35]{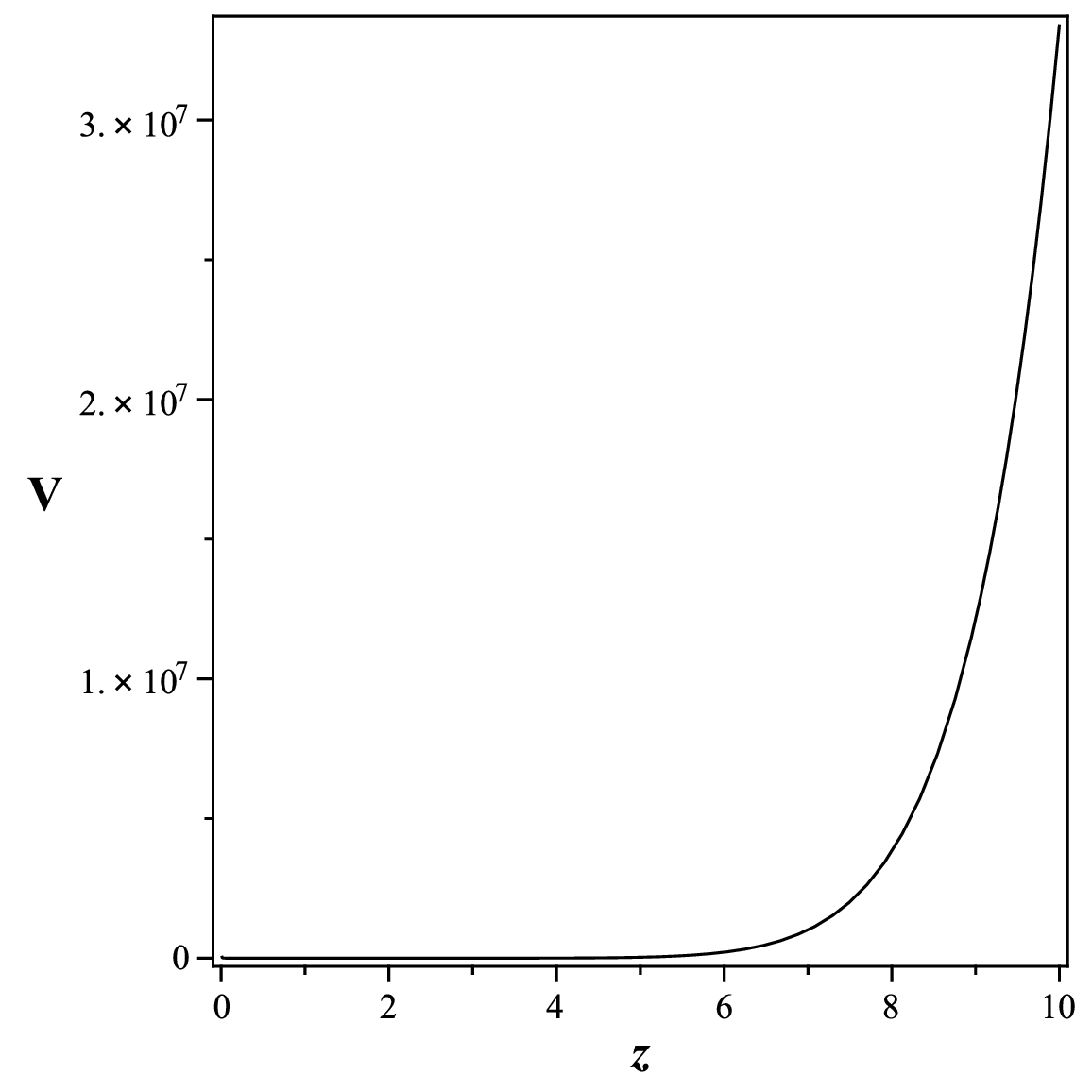}
\caption{Plot of the effective potential versus $z$ for
$q=\lambda=c=p=1$.}
\end{center}
\end{figure}

\begin{figure}[th]
\begin{center}
\includegraphics[scale=.35]{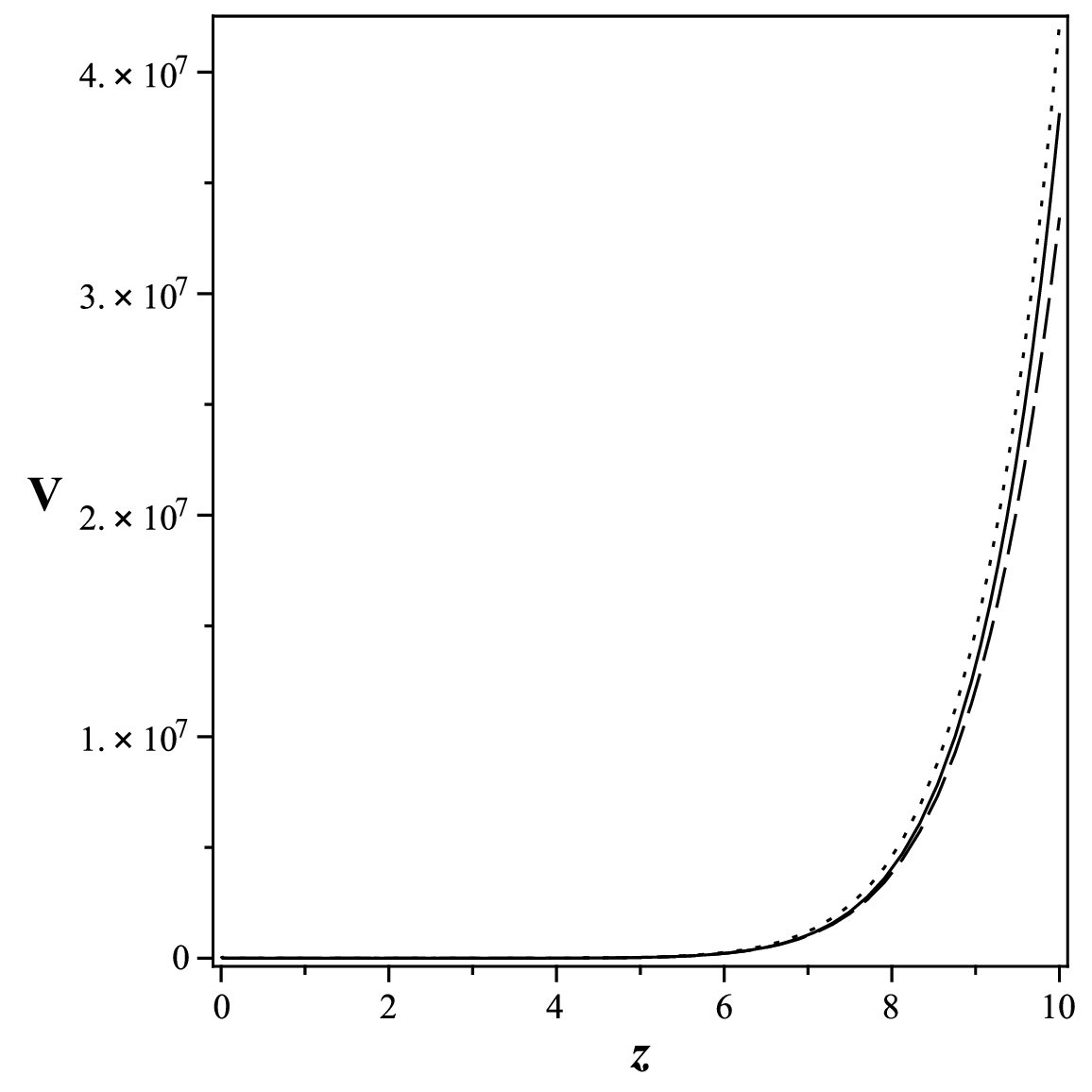}
\caption{Plot of the effective potential versus $z$ for
$q=\lambda=c_{1}=0$ (solid line), $q=c_{1}=0, \lambda=1$ (dotted
line), and $q=\lambda=0, c_{1}=1$ (dashed line).}
\end{center}
\end{figure}
\end{document}